\documentclass[a4,journal]{IEEEtran}
\usepackage{amsmath,amsfonts,amssymb}
\usepackage{algorithmic}
\usepackage{algorithm}
\usepackage{array}

\usepackage{multicol}
\usepackage{graphicx}
\usepackage{subcaption}
\usepackage{textcomp}
\usepackage{stfloats}
\usepackage{url}
\usepackage{verbatim}
\usepackage{graphicx}
\usepackage{cite}
\usepackage{cuted}
\usepackage{doi}
\usepackage{tikz}
\usetikzlibrary{shapes,arrows,positioning,decorations.pathreplacing,calc}

\begin{document}

\title{Can An Array of Magnetometers Replace a DVL --- A Cram\'{e}r-Rao Bound Analysis}

\author{Ludvig Lindström, Isaac Skog,~\IEEEmembership{Senior Member,~IEEE}, Magnus Lundberg Nordenvaad, and Andreas Gällström,~\IEEEmembership{Senior Member,~IEEE} 
\thanks{Ludvig Lindström is with the Dept. of Communication Systems, KTH Royal Institute of Technology, Sweden (email: lulinds@kth.se)

Isaac Skog is with the Dept. of Communication Systems, KTH Royal Institute of Technology, and the Div. of Underwater Technology, Swedish Defence Research Agency (FOI), Kista, Sweden (email: skog@kth.se)

Magnus Lundberg Nordenvaad is with the Div. of Underwater Technology, Swedish Defence Research Agency (FOI), Kista, Sweden (email: magnus.lundberg@foi.se)

Andreas Gällström is with SAAB Dynamics AB, Linköping, Sweden (email: andreas.gallstrom@saabgroup.com)}
\thanks{This work has been partially funded by the Vinnova, Sweden's Innovation Agency, and
Swedish Armed Forces, project 2025-03855 Electromagnetic navigation for smaller unmanned underwater vehicles.}
\thanks{This work has been submitted to the IEEE for possible publication. Copyright may be transferred without notice, after which this version may no longer be accessible.}}
\markboth{}
{Shell \MakeLowercase{\textit{et al.}}: A Sample Article Using IEEEtran.cls for IEEE Journals}
\maketitle
\begin{abstract}
Underwater vehicles rely on noisy dead-reckoning systems, such as Doppler velocity log (DVL)-aided inertial navigation systems, for positioning. However, DVLs rely on active acoustic measurements and are thus unsuitable for covert operations.  Magnetic velocity estimation using a magnetic sensor array is a potential silent alternative. In outdoor environments, spatial variations in magnetic field magnitude are small, whereas temporal variations are large, making the achievable velocity-estimation accuracy uncertain. In this paper, Cram\'{e}r-Rao lower bounds (CRLBs) are derived for magnetic velocity estimation using different sensor configurations and measurement models in one dimension, ignoring platform disturbances, to quantify the potential accuracy of magnetic sensor arrays. The analysis suggests that, when spatial variations are sufficient and temporal variations can be corrected using correction data, the resulting CRLB is comparable to that of an export-controlled DVL. If, instead, temporal variations are not corrected, the sensor noise must be reduced by around two orders of magnitude for the CRLB to reach the accuracy of an export-controlled DVL. Therefore, the results indicate that current magnetic sensor arrays may complement, rather than replace, DVLs in inertial navigation systems.

\end{abstract}


\section{Introduction}\label{sec:intro}
\IEEEPARstart{A}{utonomous} underwater vehicles (AUVs) typically cannot navigate using global navigation satellite systems underwater, since radio signals generally penetrate only a few centimeters. Therefore, AUVs usually navigate with inertial navigation systems aided by multibeam sonar and Doppler velocity logs (DVLs).
While effective, these active acoustic systems 
can reveal the vehicle's presence, making them unsuitable for covert operations.
Alternative silent sensor technologies, such as differential pressure sensor speedometers~\cite{Meurer2020} and artificial lateral line sensors~\cite{wang2026flexible}, are promising, but these technologies are still not mature.

\begin{figure}[tb!]
    \centering
     \begin{subfigure}[t]{0.48\textwidth}
        \centering
        \includegraphics[width=\linewidth]{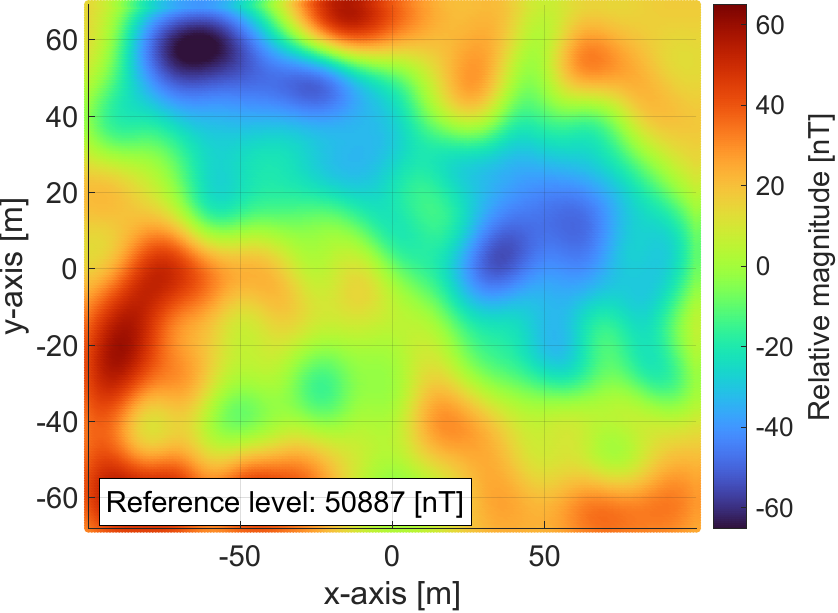}
        \caption{Spatial magnetic field magnitude variations. }
        \label{fig:field_spatial}
    \end{subfigure}\vspace{5mm}
    \begin{subfigure}[t]{0.48\textwidth}
        \centering
        \includegraphics[width=0.9\linewidth]{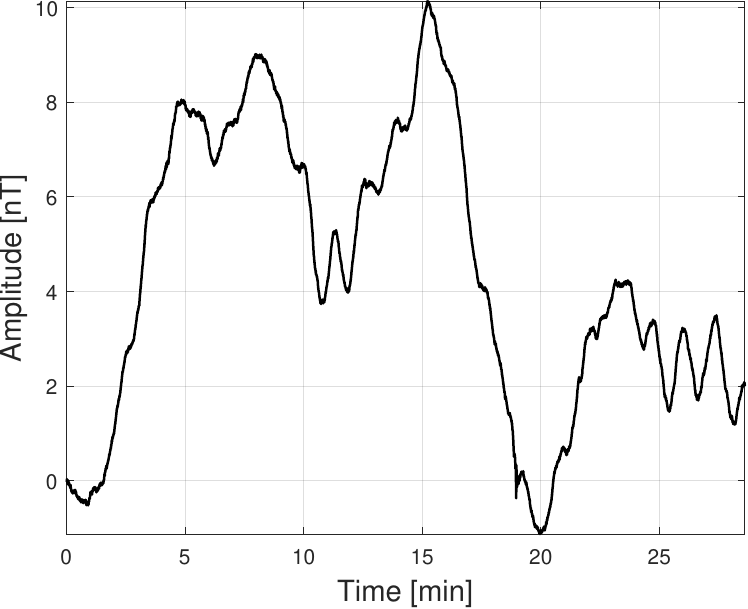}
        \caption{Temporal magnetic field magnitude variations.}
        \label{fig:field_temporal}
    \end{subfigure}
    \caption{Example of spatial variations at a height of a few meters above the sea surface at Gränsö, Sweden~\cite{skog2025underwater}. The figure also shows an example of the temporal variations in the field.}
    \label{fig:field_granso}
\end{figure}

It has been shown that spatial variations in magnetic field magnitude, as illustrated in Fig.~\ref{fig:field_spatial}, can be used for relative and absolute positioning in indoor environments~\cite{huang2023mains, edridge2025position,skog2026magnetic}. In outdoor environments, the spatial magnetic field magnitude variations, caused by magnetic material in the Earth's crust, are significantly smaller than those observed indoors~\cite{canciani2016absolute,skog2026magnetic}. Nevertheless, these spatial variations have been shown to be large enough for outdoor velocity estimation under suitable conditions~\cite{merk2025cramer}.

In addition to spatial variations, the magnetic field magnitude also varies over time, as seen in Fig.~\ref{fig:field_temporal}. These temporal variations come from multiple sources, such as solar activity and electronics in both human-made infrastructure and the sensor platform. Their magnitude depends on factors such as time of day, latitude, and the solar activity, etc.~\cite{lowrie2023earth}. In outdoor environments, the magnitude of the spatial variations caused by the Earth's crust and the temporal variations is often comparable~\cite{canciani2016absolute,lowrie2023earth,skog2026magnetic}. One way to compensate for temporal variations is to use a stationary reference station that transmits correction data, but because underwater radio communication is severely limited, this is unavailable for most AUVs. However, recently developed quantum magnetometers may be sensitive enough to both measure the small spatial variations in the outdoor magnetic field and identify and mitigate the temporal variations~\cite{Sambataro2025, krelina2021quantum}.

Therefore, it remains unclear to what degree temporal variations degrade velocity estimates from a magnetometer array. One way to quantify this is to use the Cramér-Rao lower bound (CRLB), which is the lowest variance achievable by an unbiased estimator~\cite{Kay97}. The CRLB does not describe the performance of a specific estimator. Instead, it provides a theoretical benchmark given a measurement model and sensor configuration~\cite{merk2025cramer,skog2025underwater,wang2024comparison}. To that end, this paper extends the results in~\cite{skog2025underwater} and investigates the CRLB for velocity estimation using a magnetometer array equipped with current and next-generation quantum magnetometers. The bound is compared to the accuracy of contemporary DVLs.

The main contributions of this paper are an analysis of the achievable velocity estimates 
using a uniform linear magnetometer array under different operating conditions and sensor configurations. Specifically, when the magnetic field magnitude can be adequately locally modeled as a linear function of position, temporal field variations become the performance-limiting factor. If the sensor noise is decreased, the array aperture is increased, or the total displacement over the measurement interval is increased to such an extent that higher-order models become necessary to describe the field adequately, the effect of temporal field variations can be mitigated, provided that the number of sensors is sufficient for the adopted spatial model order. In this regime, sensor noise and array geometry instead become the performance-determining factors.
The analysis is restricted to constant velocity in one dimension. It also ignores platform interference to obtain tractable analytic expressions that reveal the key performance limits and trade-offs of velocity estimation using a magnetometer array.

\section{Problem Formulation}\label{sec:problem formulation}
The problem considered in this paper is to characterize the theoretically achievable accuracy of estimating the platform velocity $v$ from a sequence $y_{0:N-1}$ of magnetometer-array measurements when the spatial field is unknown. In particular, investigating how the achievable velocity-estimation accuracy depends on the local spatial variation of the magnetic field magnitude, the temporal field variations, the sensor noise level, and the sensor-array configuration.


\subsection{General measurement model}

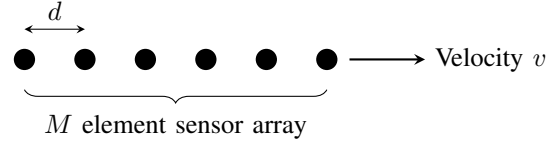
\begin{figure}[tb!]
    \centering
    \begin{tikzpicture}[>=stealth,scale=1.0]

  \def\M{6}          
  \def\spacing{0.8}  

  \foreach \m in {0,...,5} {
    \fill (\m*\spacing,0) circle (4pt);
  }

  \draw[<->] (0,0.4) -- (\spacing,0.4)
    node[midway,above] {$d$};

  \draw[decorate,decoration={brace,amplitude=5pt,mirror}]
    (0,-0.4) -- ({(\M-1)*\spacing},-0.4)
    node[midway,below=6pt] {$M \text{~element sensor array}$};

  \draw[->,thick] ({(\M-1)*\spacing+0.3},0) -- ++(1,0)
    node[right] {Velocity $v$ };

\end{tikzpicture}
    \caption{Illustration of the considered linear magnetometer sensor array and motion direction.}
    \label{fig:linear array}
\end{figure}

\begin{figure}[tb!]
        \centering
        \includegraphics[trim={3.5cm 0.5cm 3.5cm 0.5cm},width=0.9\linewidth]{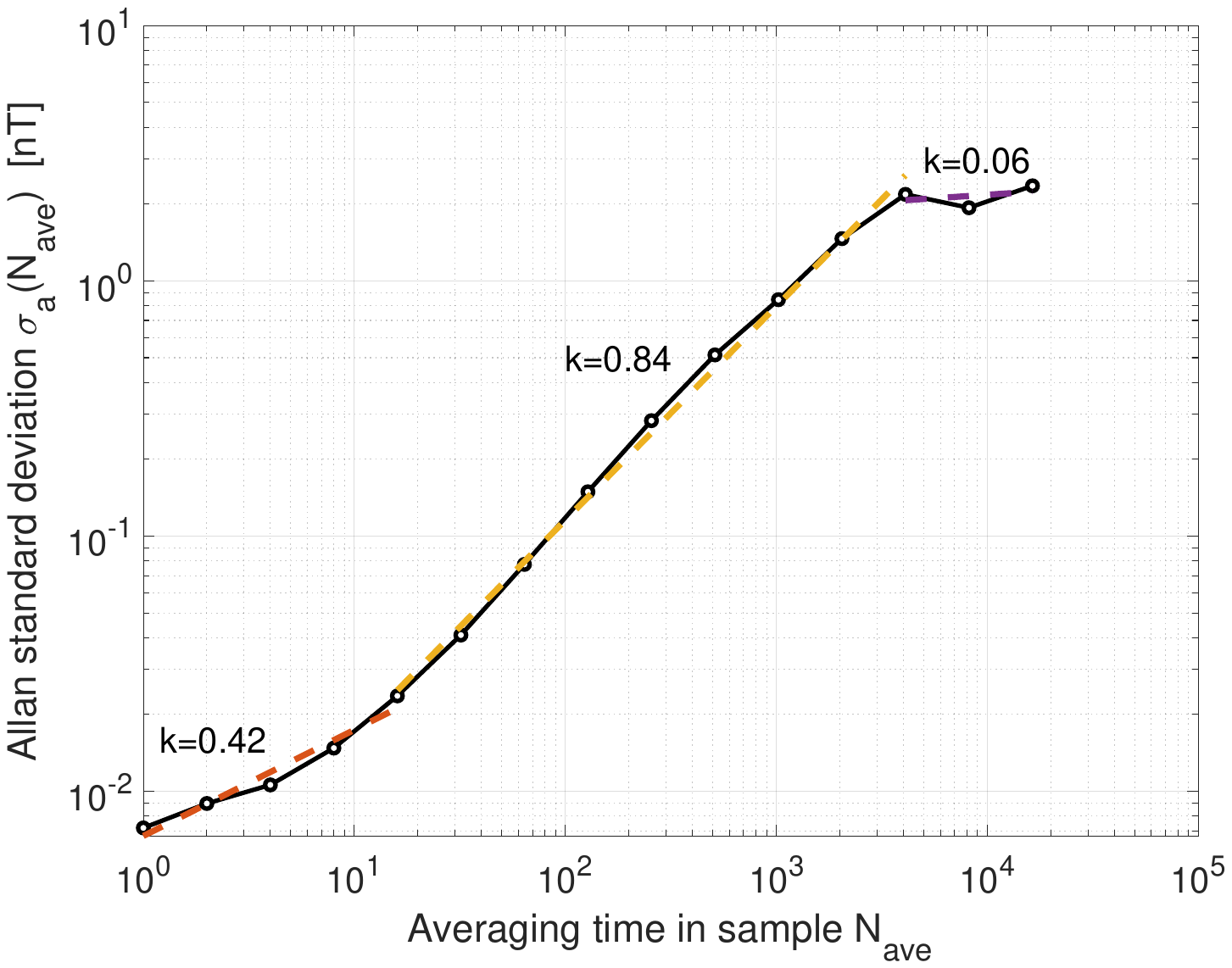}
        \caption{Allan deviation of the temporal variations at Gränsö, Sweden~\cite{skog2025underwater}, with identified slope segments and coefficients.}
        \label{fig:allan}
\end{figure}

Consider an unknown magnetic field magnitude $\mathcal{M}$ and a uniform linear array with $M$ sensors and an inter-sensor distance $d$, as illustrated in Fig.~\ref{fig:linear array}. Furthermore, assume that the array is perfectly aligned with the platform's direction of motion. Then the vector $y_n$ of magnetometer measurements from the array at time $n$ can be modeled as  
\begin{subequations}\label{eq:measurement_model}
    \begin{align}
    y_n&=\mu_n(\theta)+\eta_n 1_M + e_n,\,\,\qquad e_n\,\stackrel{\text{\tiny i.i.d.}}{\sim}\mathcal{N}(0,\sigma^2_e I_M),\\
    \eta_n&=\eta_{n-1}+T_s w_n, \,\,\,\,\, \eta_0=0, \quad w_n\stackrel{\text{\tiny i.i.d.}}{\sim}\mathcal{N}(0,\sigma^2_w).
\end{align}
\end{subequations}
Here $\mu_n(\theta)$ is a model of the magnetic field magnitude $\mathcal{M}$ across the array at time instant $n$, $T_s$ is the sampling period, \mbox{$1_n\triangleq \begin{bmatrix}
    1 & 1 & \hdots & 1
\end{bmatrix}^\top\in \mathbb{R}^n$}, and $I_{n}$ is the $n\times n$ identity matrix. The unknown model parameters $\theta$ include, for example, platform velocity and parameters describing the spatial behavior of the magnetic field magnitude.

Furthermore, $e_n$ denotes the sensor noise and is here assumed to be Gaussian. This is a conservative assumption since, among additive noise distributions with a given covariance, Gaussian noise minimizes the Fisher information about parameters entering through the measurement mean, and hence maximizes the corresponding CRLB~\cite{stoica2011gaussian}. Finally, $\eta_n$ denotes the temporal variations in the Earth's magnetic field magnitude relative to the start time $n=0$. The temporal variations are here modeled as a random walk driven by $w_n$ with variance $\sigma_w^2$. This model is motivated by the signal drift in the Allan deviation in Fig.~\ref{fig:allan}. Since the slope of the Allan deviation is roughly $0.5$, the signal drift is nearly a random walk~\cite{niu2014using}. The sensor noise and the driving noise are assumed to be independent.

Next, adopt a sample-time-major ordering of the data, and define the vectors
\begin{subequations}
\begin{align}
    \mathrm{y}=&\begin{bmatrix}
        y_0^\top & y_1^\top & \hdots &  y_{N-1}^\top
    \end{bmatrix}^\top,\\
       \mu(\theta)=&\begin{bmatrix}
        \mu_0(\theta)^\top & \mu_1(\theta)^\top & \hdots & \mu_{N-1}(\theta)^\top\end{bmatrix}^\top,
\end{align}
and the covariance matrix
\begin{equation}
    \Sigma=R_{\text{rw}} \otimes (1_M 1^\top_M)+\sigma^2_e I_{NM},
\end{equation}
where the entries of the random walk matrix are given by
\begin{equation}
    [R_{\text{rw}}]_{i,j}
    = \sigma_w^2T_s^2\min(i,j),\quad i,j\in\{0,\dots,N-1\}.
\end{equation} 
\end{subequations}
The measurements can then be modeled as
\begin{equation}
    \mathrm{y}\sim\mathcal{N}(\mu(\theta),\Sigma). 
\end{equation}

\subsection{Spatial field models}

The magnetic field magnitude, $\mathcal{M}$, is generally a smooth function. Therefore, it can be represented using a Taylor expansion around $x_0$, i.e.,
\begin{subequations}
\begin{equation}
    \mathcal{M}(x)=\gamma+\beta(x-x_0)+\alpha(x-x_0)^2+\mathcal{O}((x-x_0)^3)
\end{equation}
where for brevity \begin{align}
    \gamma=&\left.\mathcal{M}\right|_{x=x_0},\\ \beta=&\left.\frac{\partial\mathcal{M}}{\partial x}\right|_{x=x_0},\\
    \alpha=&\frac{1}{2}\left.\frac{\partial^2\mathcal{M}}{\partial x^2}\right|_{x=x_0},
\end{align}
\end{subequations}
and $\mathcal{O}$ denotes the big O notation.

A rough indication of when the second-order spatial variation is small relative to the sensor noise is if
\begin{equation}
    |\alpha| (L+(N-1)|\Delta|)^2\ll \sigma_e,
\end{equation}
where $\Delta\triangleq T_sv$ the displacement between each measurement, $L\triangleq (M-1)d$ is length of the array, and $N$ is the number of measurements.
Therefore, depending on the region of interest, the size of the array, and the approximated magnetic field magnitude model $\mu(\theta)$
may require different orders to describe the true field $\mathcal{M}(x)$ sufficiently accurately.

\subsection{Classical and quantum noise levels}

Quantum magnetometers, including optically pumped and nitrogen-vacancy center magnetometers, are becoming increasingly available outside of the laboratory~\cite{degen2017quantum, fabricant2023build, Sambataro2025}. 
For these sensors, the total measurement noise can broadly be separated into quantum and classical readout terms~\cite{degen2017quantum}.
Current sensors have a total noise level of approximately $1$~pT/$\sqrt{\text{Hz}}$, while the quantum noise is anticipated to reach below $1$~fT/$\sqrt{\text{Hz}}$~\cite{Sambataro2025}. In the coming years, classical readout noise is expected to decrease significantly as the technology develops~\cite{degen2017quantum}.




\section{Theoretical Performance Bounds}\label{sec:crlb}

In this section, five CRLBs for velocity estimation accuracy will be studied. They are based upon different assumptions regarding the spatial and temporal behavior of the magnetic field magnitude.  

The general Gaussian CRLB for $\theta$, where the covariance $\Sigma$ is independent of $\theta$, is given by~\cite{Kay97}
\begin{equation}\label{eq:kay_crlb}
    \text{Cov}(\hat{\theta})\succeq \mathcal{I}^{-1}_\theta, \qquad \mathcal{I}_\theta=\frac{\partial \mu(\theta)^\top}{\partial\theta} \Sigma^{-1}\frac{\partial \mu(\theta)}{\partial\theta},
\end{equation}
where $A\succeq B$ denotes that $A-B$ is positive semi-definite.
Further, if the magnetic field magnitude can be modeled as

\begin{equation}\label{eq:crlb_projection_condition}
    \mu(\theta)=A(\theta_{nl})\theta_{l},\qquad \text{dim}(\theta_{nl})=1,
\end{equation}
then it holds that~\cite{stoica1989music}
\begin{subequations}\label{eq:crlb_projection}
\begin{equation}
    \text{var}(\hat{\theta}_{nl})\geq \frac{1}{g(\theta_{nl})^{\top} \Sigma^{-1} P_{\perp} g(\theta_{nl})}
\end{equation}
where 
\begin{equation}
    g(\theta_{nl}) =\frac{\partial}{\partial \theta_{nl}}A(\theta_{nl})\theta_l    
\end{equation}
and the projection matrix is given by
\begin{equation}
P_{\perp}= I- A(\theta_{nl})
\bigl(A(\theta_{nl})^{\top} \Sigma^{-1} A(\theta_{nl})\bigr)^{-1}
A(\theta_{nl})^{\top} \Sigma^{-1}.  
\end{equation}
\end{subequations}

\subsection{Unknown deterministic signal and no temporal variations}\label{sec:Unknown deterministic signal and no temporal variations}
Assuming that the spatial field is deterministic and there are no temporal variations, the velocity can be estimated by correlating the measurements across the array. To that end, let $\breve{\mu}(\theta)$ and $\breve{\Sigma}$ denote the mean and covariance of the data when a sensor-major sample ordering is used. Further, let $\theta_{nl}=\tau$, where $\tau\triangleq d/v$. That is, $\tau$ is the time difference between when two neighboring sensors see the same spatial features. Moreover, let  
\begin{equation}
\theta_l=\begin{bmatrix}
    s_0 & s_1 & \hdots & s_{N-1}
\end{bmatrix}^\top,\quad s_n\triangleq\mathcal{M}(x_n),    
\end{equation}
denote the magnetic field magnitude along the array trajectory given by the positions $x_n=vnT_s$, $n\in\{0,1,\ldots, N-1\}$.  Then, when $N\rightarrow\infty$, the field  $\breve{\mu}(\theta)$ can be described as  
\begin{subequations}    
\begin{equation}
    \breve{\mu}(\theta)=A(\theta_{nl})\theta_l 
\end{equation}
where
\begin{equation}
    A(\tau)=\begin{bmatrix}
        D(0)\\
        D(\tau)\\
        \vdots\\
        D((M-1)\tau)
    \end{bmatrix}.
\end{equation}
\end{subequations}
Here $D(\tau)$ is an ideal fractional-delay operator, such that $D(\tau)^\top D(\tau)=I$ for all delays $\tau$. Note that such an ideal fractional delay operator requires an infinitely long signal, and any real fractional delay operator will be an approximation thereof.  

Next, assuming there are no temporal variations, i.e. \mbox{$\breve{\Sigma}=\sigma^2_e I_{NM}$}, and using the CRLB expression in \eqref{eq:crlb_projection_condition}, it holds that, 
\begin{equation}\label{eq:delay_crlb}
  \text{var}(\hat{\tau})\gtrsim\frac{12\sigma^2_e}{M(M^2-1)\sum_{n=0}^{N-1} \dot{s}^2_n},\qquad N\gg1
\end{equation}
where $\dot{s}_{n}$ is the time derivative of $s$ at time $n$. Here, $\gtrsim$ denotes an approximate lower bound, which asymptotically converges to the actual lower bound as $N\rightarrow \infty$.  

Assume that the field is linear, i.e.,
\begin{equation}
    \mathcal{M}(x)=\gamma+\beta x.
\end{equation}
Then, using the CRLB variable transformation rule~\cite{Kay97}, it holds that 
\begin{equation}\label{eq:bound1}
\boxed{
    \text{var}(\hat{v})\gtrsim \frac{12\,\sigma^2_e v^2}{M(M^2-1)N\beta^2d^2},\qquad  N\gg 1.}
\end{equation}
Note that, apart from assuming a deterministic field and no temporal variations, this bound does not require any assumptions about the prior knowledge of the spatial behavior of the magnetic field magnitude.

\subsection{Linear field model and no temporal variations}

Assume that it is known that the magnetic field magnitude can be approximated locally as linear, then,
\begin{subequations}\label{eq:first_order_eq}
\begin{equation}
\begin{split}
\mu_n(\theta) =& \gamma\,1_M+\beta\left(v\,n\,T_s\,1_M + d\,u_M\right),
\end{split}
\end{equation}
where
\begin{equation}
u_M=\begin{bmatrix}
        0 \\ 1 \\\vdots \\ M-1
    \end{bmatrix}-\frac{(M-1)}{2}1_M\quad\text{and}\quad \theta=\begin{bmatrix}
    v \\  \gamma \\ \beta
\end{bmatrix}.
\end{equation}
\end{subequations}
Note that this approximated field fulfills the conditions in \eqref{eq:crlb_projection_condition}, with $\theta_{nl}=v$ and $\theta_l=\begin{bmatrix}
    \gamma & \beta
\end{bmatrix}^\top$.
Also note that the number of unknown parameters is reduced from $N+1$ to $3$, compared to the general deterministic magnetic field magnitude in Section~\ref{sec:Unknown deterministic signal and no temporal variations}. 

%
Thus, it can be shown that if $M\geq 2$, the CRLB for the velocity is given by
\begin{equation}\label{eq:bound2}
\boxed{
\text{var}(\hat{v})\geq\frac{12\,\sigma^2_e v^2}{M(M^2-1)N\beta^2d^2}+\frac{12\,\sigma^2_e}{MN(N^2-1)\beta^2T_s^2}}
\end{equation}
This bound converges to the bound in \eqref{eq:bound1} as $N\rightarrow \infty$. The bound may be rewritten to 
\begin{equation}\label{eq:crlb speed}
\text{var}(\hat{v})\geq \frac{12\,\sigma_e^2}{MN\beta^2 T_s^2}\left(\frac{(M-1)}{(M+1)}\frac{\Delta^2}{L^2}+\frac{1}{(N^2-1)}\right)
\end{equation}
This shows that it is beneficial to keep the displacement $\Delta$ small relative to the array length $L$ to ensure significant overlap between measurement points. Note that, as the field can only be well approximated as linear over a limited distance, the bound in~\eqref{eq:bound2} is generally more useful than the bound in~\eqref{eq:bound1} when evaluating the expected performance.

\subsection{Linear field model and temporal variations}
Assume that it is known that the first-order approximation of the magnetic field magnitude also has temporal variations, i.e., $\sigma_w>0$. Then deriving an analytically tractable exact expression for the CRLB for the velocity $\hat{v}$ is difficult. However, an approximate bound has to fulfill the following conditions if $M\geq2$,
\begin{subequations}
\begin{align}
    \sigma_w\rightarrow&\;0\implies  \text{Bound in }\eqref{eq:bound2}\\
    \sigma_e\rightarrow&\;0\implies \text{var}(\hat{v})\geq\frac{\sigma_w^2}{(N-1)\beta^2}.
\end{align}
\end{subequations}
The second case follows from the fact that the slope $\beta$ and offset $\gamma$ can be perfectly estimated from the first measurement if $M\geq2$. Hence, estimating the velocity $v$ becomes equivalent to estimating a constant in noise.

An approximate bound, motivated by small spectral overlap of the two noise terms,  is therefore given by the sum of the bounds in the two extreme cases. Here the sensor noise has a white spectrum, and the temporal variance has a low-frequency-dominant spectrum.
%
%
Therefore, an approximate bound is given by
\begin{equation}\label{eq:bound3}
\boxed{
\begin{aligned}
\text{var}(\hat{v})\gtrsim& \frac{12\,\sigma^2_e v^2}{M(M^2-1)N\beta^2d^2}+\frac{12\,\sigma^2_e}{MN(N^2-1)\beta^2T_s^2}\\
&+\frac{\sigma^2_w}{(N-1)\beta^2}.
\end{aligned}}
\end{equation}

\subsection{Second-order field model and no temporal variations}
Assume instead that it is known that the magnetic field magnitude can be approximated locally as a second-order polynomial, i.e.,
\begin{subequations}\label{eq:second_order_eq}\begin{equation}
\begin{split}
\mu_n(\theta) =& \gamma\, 1_M+\beta\left(v\,n\,T_s\,1_M + d\,u_M\right)\\
+& \alpha\left(v\,n\,T_s\, 1_M+ d\,u_M \right)^{\circ2},\end{split}     
\end{equation}
where
\begin{equation}
 \theta=\begin{bmatrix}
    v & \gamma & \beta & \alpha
\end{bmatrix}^\top,
\end{equation}
\end{subequations}
and $(\cdot)^{\circ2}$ is the element-wise square operation. Note that this measurement model also fulfills \eqref{eq:crlb_projection_condition}.
Consider the case where the temporal variations $\sigma_w=0$ and the number of sensors in the array $M\geq 2$. Then the CRLB is
\begin{equation}\label{eq:bound4}
    \boxed{
    \begin{aligned}
    \text{var}(\hat{v})\geq&\frac{12\, \sigma_e^2}{MN}\left(\frac{v^2}{(M^2-1)d^2}+\frac{1}{(N^2-1)T_s^2}\right)\\
    &\times\biggl[\beta^2+2\,\alpha\,\beta\, v \,T_s(N-1)\\
    &+\frac{2\,\alpha^2}{3}v^2 T_s^2(N-1)(2N-1)\\
    &+\frac{\alpha^2}{3}d^2(M^2-1)-\alpha^2A \biggr]^{-1},
    \end{aligned}}
\end{equation}
where for $M\geq 2$, \begin{equation}
\begin{split}
    A=&(7M^2N^2-3N^2-3M^2-13)\\
    &\times\left(\frac{(N^2-1)}{d^2}+\frac{(M^2-1)}{v^2T_s^2}\right)\\
    &\times \biggl[\frac{(N^2-1)(N^4-4)}{5d^4}+\frac{(M^2-1)(M^2-4)}{5v^4T_s^4}\\
    &+\frac{(M^2-1)(N^2-1)}{v^2T_s^2d^2}\biggr]^{-1}\geq 0.
\end{split}
\end{equation}
A non-tight lower bound of the estimation variance can be written as, 
\begin{equation}\label{eq:bound4_non_tight}
    \begin{split}
    \text{var}(\hat{v})\geq&\frac{12\, \sigma_e^2}{\beta^2T_s^2MN}\left(\frac{(M-1)}{(M+1)}\frac{\Delta^2}{L^2}+\frac{1}{(N^2-1)}\right)\\ 
    &\times\biggl[1+2\frac{\alpha}{\beta} \Delta(N-1)\\
    &+ \frac{2\,\alpha^2\Delta^2}{3\beta^2}(N-1)(2N-1)\\
    &+\frac{\alpha^2L^2}{3\beta^2}\frac{(M+1)}{(M-1)}\biggr]^{-1}.
    \end{split}
\end{equation}
Note that when $\alpha=0$, the bound coincides with that in \eqref{eq:bound2}, and since $\alpha/\beta$ is typically small, the CRLB with the second-order model is in most cases nearly equivalent to the CRLB using the first-order one. However, when $\alpha$ is large, the CRLB decreases much more rapidly with the number of measurements and the sensor array length. As $N\rightarrow \infty$, the bound decreases at a rate of $1/N^3$ using the second-order model, compared to $1/N$ using the first-order model. Similarly, as $L\rightarrow\infty$, it decreases at a rate of $1/L^2$ for the second-order model, compared to converging to a constant for the first-order model.



\subsection{Second-order field model and temporal variations}\label{sec:Second-order field model and temporal variations}

 Finally, assume that it is known that the magnetic field magnitude is a second-order polynomial with temporal variations. The exact CRLB is too complex to be analytically tractable. Instead, an approximation is considered. The approximate bound has to fulfill the following two conditions if $M\geq 3$,
\begin{subequations}\label{eq:condition_alpha}
\begin{align}
    \sigma_w\rightarrow&\;0\implies  \text{Bound in }\eqref{eq:bound4}\\
    \sigma_e\rightarrow&\;0\implies \text{var}(\hat{v})\geq0.\label{eq:condition_alpha_II}
\end{align}
\end{subequations}
The second case follows from the fact that then $\alpha$, $\beta$, and $\gamma$ can be determined from the first measurement if $M\geq3$. Hence $v$ can be determined exactly from the second measurement. When instead the measurement noise is much smaller than the temporal variations, that is, $\sigma^2_e\ll \sigma^2_w T_s^2$
 and $M\geq 3$
\begin{equation}\label{eq:bound5}
\boxed{
\begin{aligned}
     \text{var}(\hat{v})\gtrsim&\frac{180\,\sigma_e^2v^2}{M(M^2-1)(M^2-4)N\alpha^2d^4 }\\
     &+\frac{36\,\sigma_e^2}{M(M^2-1)N(N^2-1)\alpha^2d^2T_s^2}
     \end{aligned}
     }
 \end{equation}
See Appendix for derivation.  This bound can be rewritten to
\begin{equation}\label{eq:bound6}
\begin{split}
    \text{var}(\hat{v})\gtrsim&\frac{36\, \sigma_e^2 }{\alpha^2T_s^2L^2MN}\frac{(M-1)}{(M+1)
    }\\
    &\times\left(\frac{5(M-1)^2}{(M^2-4)}\frac{\Delta^2}{L^2}+\frac{1}{(N^2-1)}\right).
    \end{split}
\end{equation}
Note that the bound does not depend on the temporal variations even though it is the dominating term in the measurement covariance. Furthermore, sensor array length $L$ affects the CRLB more than in the first-order cases. That is, as $L\rightarrow \infty$, the CRLB decreases as $1/L^2$ compared to converging to a constant.

Also note that the first term of the bound is the CRLB of $\alpha$ for $N$ uncorrelated measurements, and the second term is the CRLB of $v$ given a known $\alpha$ and unknown $\beta$ and $\gamma$. Considering the conditions in~\eqref{eq:condition_alpha}, the exact CRLB for the second-order magnetic field magnitude approximation with temporal variations must change regimes with $\sigma_e^2/\sigma^2_w T_s^2$.

  \subsection{Higher-order approximations}
In this paper, the derived bounds assume that the magnetic field magnitude is sufficiently well described by first- and second-order polynomials, respectively. If the approximation error exceeds the sensor noise, this can introduce significant bias in the estimated velocity. Therefore, for fast-moving platforms, large sensor arrays, and sensors with low sensor noise, a higher-order approximation is preferable. 


Furthermore, assuming temporal variation but no sensor noise, using the same argument as in Section~\ref{sec:Second-order field model and temporal variations}, if the polynomial order of the approximation $P\geq 2$ and the number of sensors $M\geq P+1$, then the platform velocity can be determined exactly. This suggests that as the sensor measurement noise decreases, $\sigma^2_e$, and enough sensors are used, the temporal variations can be mitigated, meaning that the CRLB is not dependent on $\sigma^2_w$ for $ \sigma_e^2 \ll \sigma_w^2 T_s^2$.

\section{Analysis and Discussion}\label{sec:analyse}

\begin{figure}[tb!]
    \centering
        \includegraphics[trim={5cm 0.5cm 5cm 0.5cm},width=0.79\linewidth]{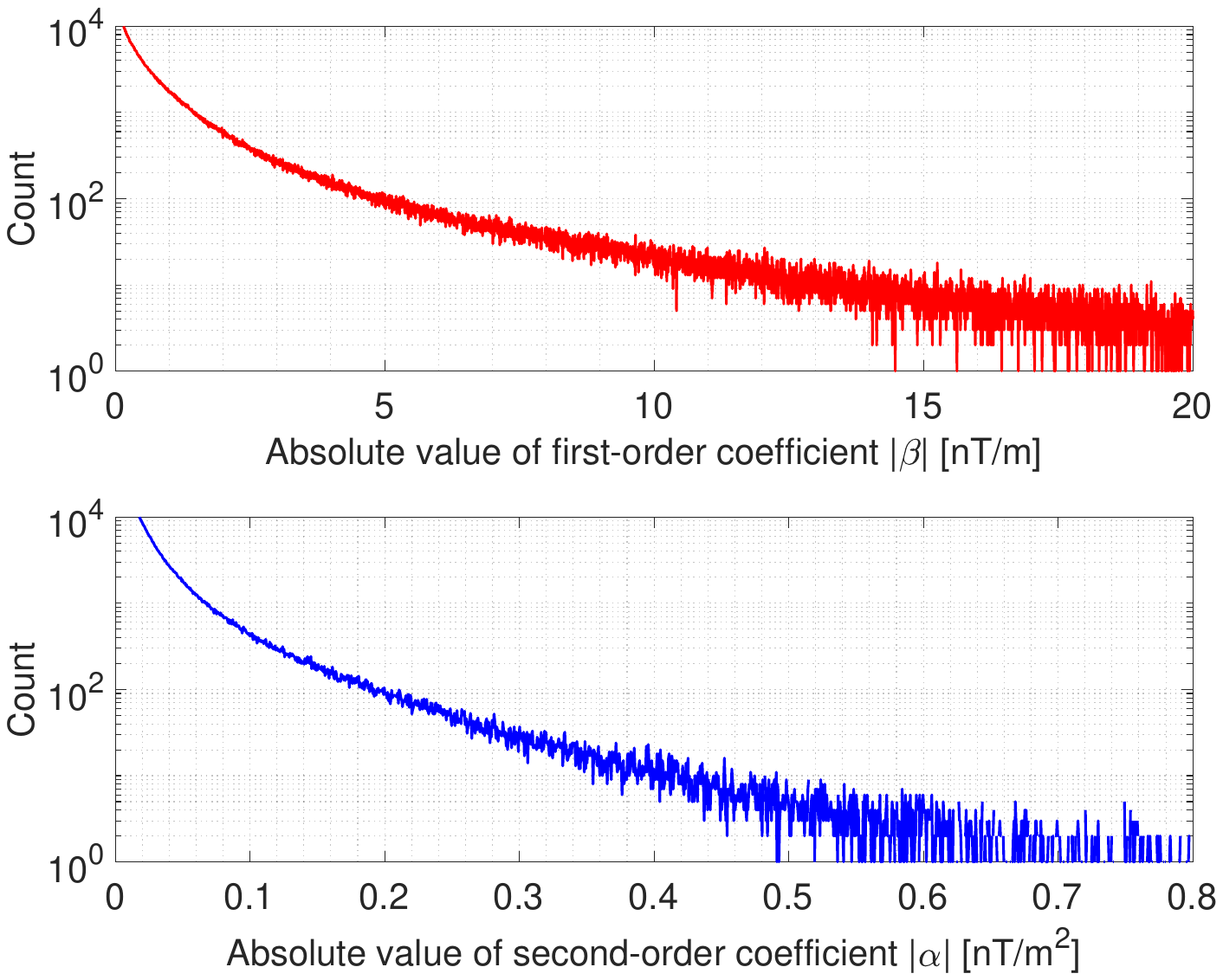}
    \caption{The counts of first- and second-order coefficients using a second-order fit over magnetometer data collected across Sweden at an altitude of around $60$~m and transformed to an altitude of $0$~m using a downward continuation.}
    \label{fig:hist_count}
\end{figure}

Next, we illustrate and discuss the analytical and numerical bounds using representative system parameters. The parameters for the first- and second-order magnetic field magnitude are taken from magnetometer data collected across Sweden by the Geological Survey of Sweden (SGU)~\cite{SGU_magnetic_aerial}. We also compare the bounds to the typical accuracy of license-free and export-controlled DVLs.

\subsection{Sensor noise and random walk noise levels}

A commercially available optically pumped total-field magnetometer has a typical noise level of $3$~pT/$\sqrt{\text{Hz}}$~\cite{quspin_qtfm_gen2}. Assuming that the sample frequency is $25$~$\text{Hz}$ and a perfect anti-aliasing low-pass filter is used, the measurement noise standard deviation is $\sigma_e\approx 0.011$~nT. 

Furthermore, the Allan deviation plot in Fig.~\ref{fig:allan} suggests that the standard deviation of the driving noise is
\begin{equation}
\begin{split}
    \sigma_w\approx&\sqrt{\frac{3}{N_{\text{\tiny ave}}}}\frac{\sigma_a(N_{\text{\tiny ave}})}{T_s}\approx\sqrt{\frac{3}{16}}\frac{0.0236}{1/25}\; \text{nT/s}\\
    \approx&0.256\; \text{nT/s}.
\end{split}
\end{equation}
Here $N_{\text{\tiny ave}}$ is the number of averaging samples~\cite{niu2014using,el2008analysis}.

\subsection{Estimation of spatial field coefficients}

The first- and second-order coefficients, $\beta$ and $\alpha$, are estimated using magnetometer data collected at around $60$~m altitude from various locations in Sweden~\cite{SGU_magnetic_aerial}. These aerial data are transformed using downward continuation with a low-pass frequency filter to approximate the magnetic field magnitude at the surface~\cite{wang2023comparison}. 

The empirical distribution of the first- and second-order coefficients obtained from a sliding second-order model fit to the downward continuation transformed data is presented in Fig.~\ref{fig:hist_count}
The mean absolute values of the coefficients are
\mbox{$|\beta|\approx1.127$~nT/m} and 
\mbox{$|\alpha|\approx0.021$~nT/m$^2$}. 
Since short-wavelength spatial variations are strongly attenuated with altitude and cannot be reliably recovered by downward continuation, the true spatial coefficients are expected to be larger than those estimated from the data.


\begin{table}[tb!]
    \centering
    \caption{Parameter values used in the numerical evaluation of the CRLB expressions.}
    \begin{tabular}{ |c|c|}
        \hline
        Platform velocity, $v$ & $3$ m/s  \\
        \hline
        Sampling period, $T_s$ &  $1/25$ s\\
        \hline
        Sensor noise, $\sigma_e$ & $0.011$ nT\\
        \hline
        Random walk process noise, $\sigma_w$ & $0.256$ nT/s\\
        \hline
        Array length, $L$ & $2$ m\\
        \hline
        Number of sensors, $M$ & $3$\\
        \hline
        First-order coefficient, $\beta$ & $1$ nT/m\\
        \hline
        Second-order coefficient, $\alpha$ & $0.02$ nT/m$^2$\\
        \hline
    \end{tabular}
    \label{tab:platform_values}
\end{table}

\begin{figure*}[tb!]
    \centering
    \begin{subfigure}[t]{0.4\textwidth}
        \centering
        \includegraphics[trim={4.5cm 1.7cm 4.5cm 0.5cm},width=\linewidth]{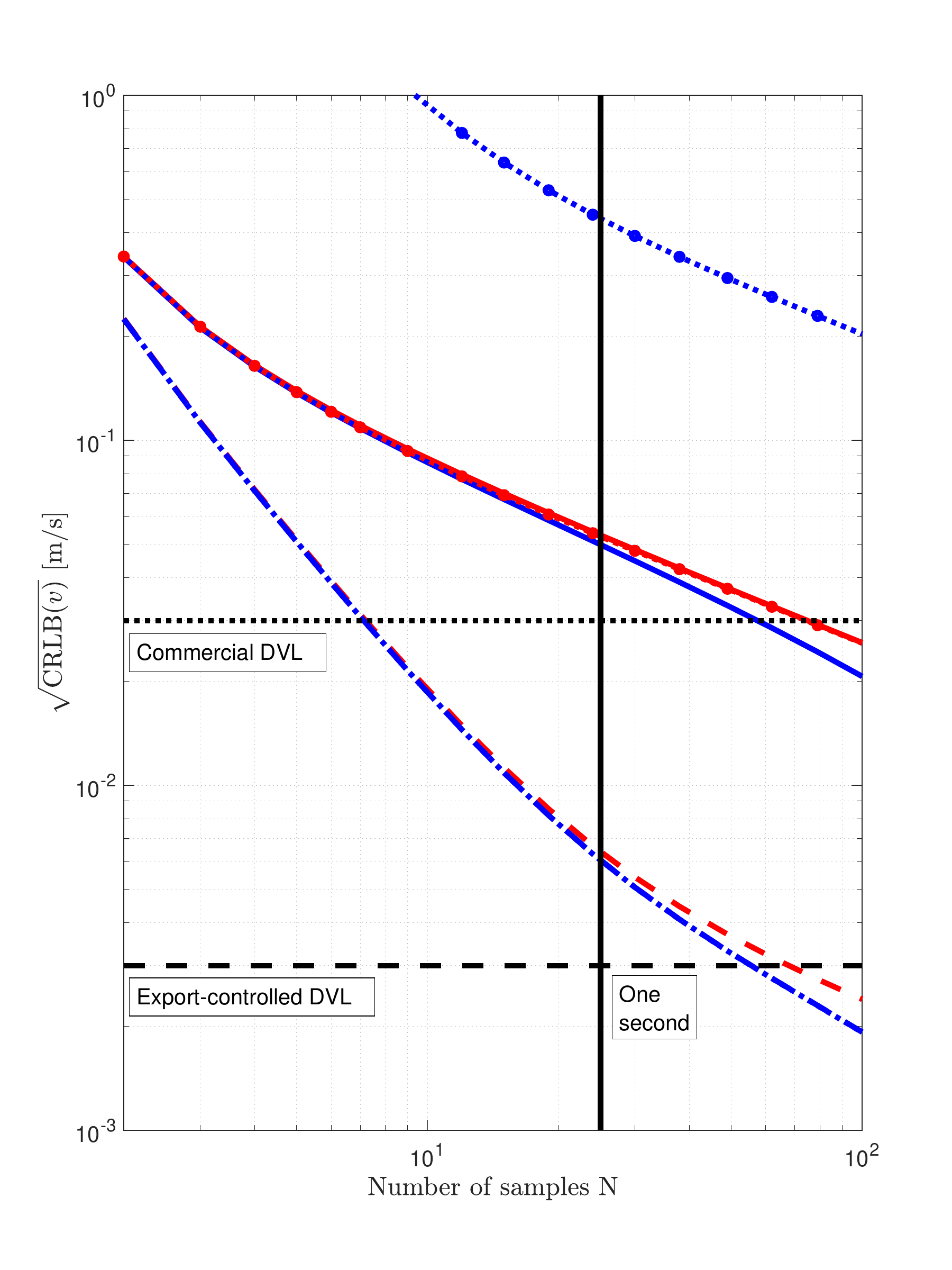}
        \caption{CRLB versus number of samples.}
        \label{fig:speed versus samples}
    \end{subfigure}
    \hspace{2cm}
    \begin{subfigure}[t]{0.4\textwidth}
        \centering
        \includegraphics[trim={4.5cm 1.7cm 4.5cm 0.5cm},width=\linewidth]{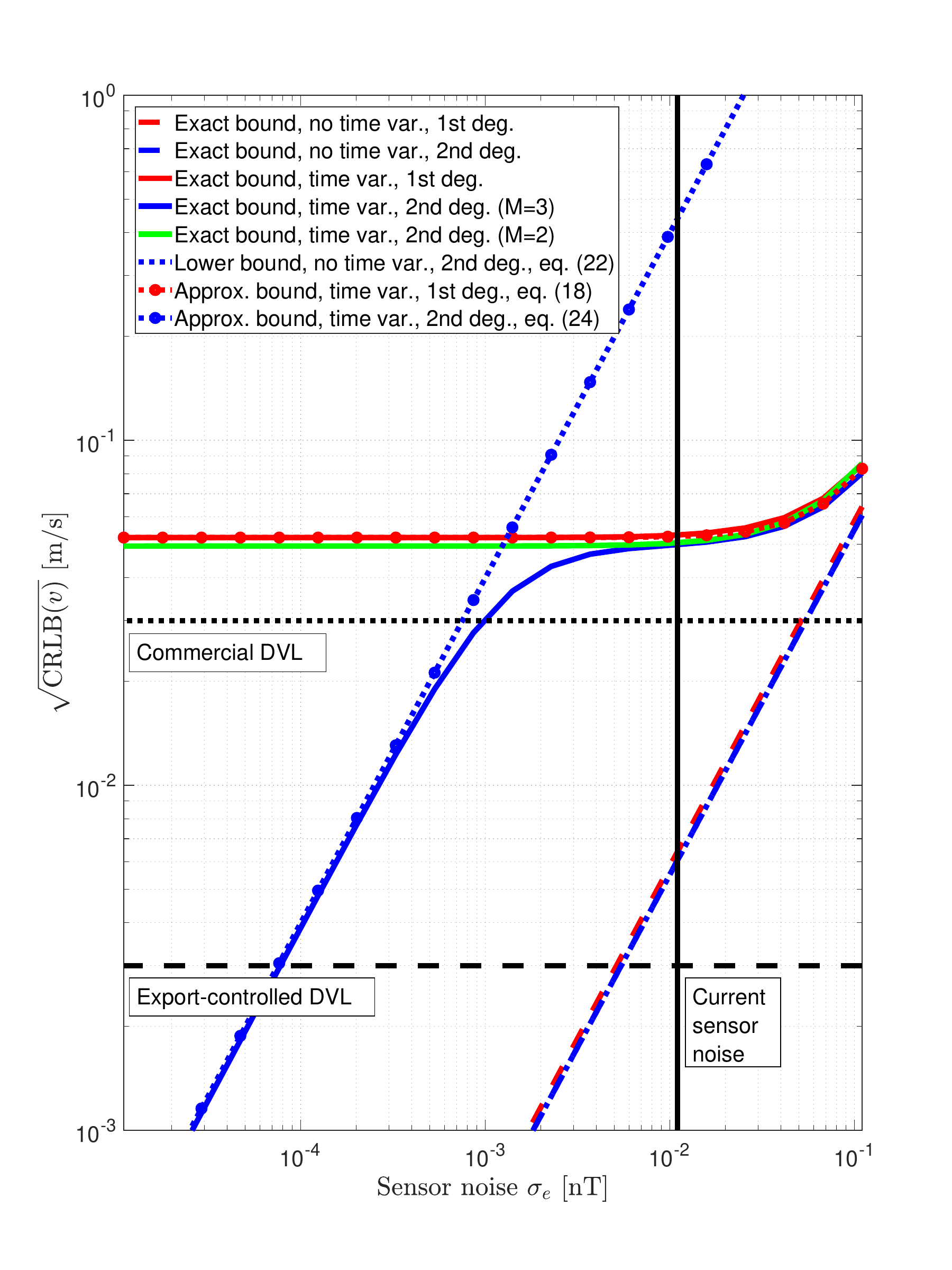}
        \caption{CRLB versus sensor noise for $N=25$ measurements.}
        \label{fig:speed versus noise}
    \end{subfigure}
    \caption{Exact and approximate CRLB of the velocity for the different model and noise assumptions. Here the array is assumed to be $2$ m, equipped with $3$ evenly spaced sensors, and moving at $3$ m/s.}
    \label{fig:crlb speed}
\end{figure*}

Based on the numbers identified from the data, the parameters used to evaluate the analytical and exact, numerically evaluated, bounds are
summarized in Table~\ref{tab:platform_values}. The resulting bounds are
shown in Fig.~\ref{fig:crlb speed}, together with the expected accuracy of commercial and export-controlled DVLs of $1\%$ and $0.1\%$ of
the velocity~\cite{nortek_dvl_manual}.

\subsection{Accuracy using a first-order model}

In Fig.~\ref{fig:speed versus samples}, the bounds in~\eqref{eq:bound2} and~\eqref{eq:bound3}, together with the exact CRLB, evaluated numerically for the conditions in~\eqref{eq:bound3}, are shown as a function of the number of samples $N$. From the figure, the approximated CRLB in~\eqref{eq:bound3} and the exact CRLB agree well for the parameter values studied.  Furthermore, for a first-order model of the magnetic field magnitude, if temporal variations can be mitigated with correction data, the typical spatial magnetic variations are large enough to estimate platform velocity with accuracy comparable to a commercial DVL after only $0.3$ seconds. On the other hand, if temporal variations cannot be mitigated, the accuracy using a first-order model is considerably worse and reaches a commercial DVL's accuracy only after around $3$~seconds.

In Fig.~\ref{fig:speed versus noise}, the bounds in~\eqref{eq:bound2} and~\eqref{eq:bound3} as well as the numerically evaluated exact CRLB, for the conditions in~\eqref{eq:bound3}, are shown as a function of the sensor noise.  Again, the approximated CRLB in~\eqref{eq:bound3} coincides well with the exact CRLB. If correction data can correct temporal variations, even a small improvement in sensor noise could make magnetometer sensor-array accuracy match an export-controlled DVL. The figure also shows that even with sensor noise much lower than that of current sensors, variance cannot be reduced below \mbox{$\sigma_w^2/((N-1)\beta^2)$} if temporal variations cannot be mitigated with correction data. This means that, for a first-order approximation of the magnetic-field magnitude with typical spatial variations and without correcting temporal variations with correction data, a magnetometer array's accuracy is generally worse than that of a commercial DVL and has little room for improvement.

\subsection{Accuracy using a second-order model}

Fig.~\ref{fig:speed versus samples} also shows the bounds in~\eqref{eq:bound4_non_tight} and~\eqref{eq:bound5} as well as the respective exact CRLB, evaluated numerically, as a function of the number of samples $N$. The figure suggests that the non-tight bound in~\eqref{eq:bound4_non_tight} agrees well with the exact CRLB, while the approximate CRLB in~\eqref{eq:bound5} does not coincide with the exact CRLB for the studied parameters.
Furthermore, for typical spatial and temporal variations and the current sensor noise level, the second-order model does not seem to significantly improve accuracy compared to a first-order model, since the second-order spatial variations are too small and the sensor noise too large.

In Fig.~\ref{fig:speed versus noise}, the bounds in~\eqref{eq:bound4} and~\eqref{eq:bound5}, as well as the respective numerically evaluated exact CRLB, are shown as a function of the sensor noise. From the figure, the lower bound in~\eqref{eq:bound4_non_tight} coincides well with the exact CRLB. In contrast, the approximate CRLB in~\eqref{eq:bound5} approaches the exact CRLB as the sensor noise decreases, increasingly satisfying the condition $\sigma_e^2\ll\sigma_w^2T_s^2$. The figure also shows that as sensor noise decreases, the second-order model, where temporal noise is not mitigated, diverges from the first-order model. This means that for a representative magnetic field magnitude, a $2$~m magnetometer sensor array with $3$ sensors can achieve an accuracy comparable to commercial and export-controlled DVLs if the sensor noise is reduced by approximately one and two orders of magnitude, respectively. Fig.~\ref{fig:speed versus noise} also shows the numerically evaluated exact CRLB for the conditions in~\eqref{eq:bound5} but with only two sensors. This bound has the same limiting behavior as the first-order model for the parameters considered.

\subsection{Effect of array length}


\begin{table}[tb!]
    \centering
    \caption{Sensor array length needed to reach the accuracy of commercial and export-controlled DVLs, numerically evaluated with exact CRLB for the conditions in \eqref{eq:bound5}, using the parameters in Table \ref{tab:platform_values}.}
    \begin{tabular}{ |c|c|c|}
        \hline
        Noise level & Commercial & Export-controlled  \\
        \hline
        $\sigma_e=0.011$~nT & $8.24$~m &  $88.9$~m\\
        \hline
        $\sigma_e=0.0011$~nT & $2.13$~m & $10.4$~m\\
        \hline
        $\sigma_e=0.00011$~nT & $0.661$~m & $2.41$~m\\
        \hline
    \end{tabular}
    \label{tab:len_rel}
\end{table}

The length of the array greatly affects the CRLB when considering the condition in \eqref{eq:bound6}. Table~\ref {tab:len_rel} reports the sensor-array lengths needed for the CRLB to reach the accuracy of commercial and export-controlled DVLs using current and upcoming generations of magnetometers, given the parameters in Table~\ref{tab:platform_values}. Since the second-order model is only valid locally, as indicated by the field in Fig.~\ref{fig:field_spatial}, if the sensor array is too large, the approximation error might exceed the sensor noise. 

\subsection{Practical challenges}

The presented bounds assume that only sensor noise and temporal variations affect the measurements. However, when implementing a magnetic sensor array on an AUV, one would face many practical challenges, one major being platform disturbances. These disturbances include static sources such as permanent and induced magnetization and eddy currents, as well as time-varying sources such as electronics and motors. The magnetic field magnitude produced by static sources can be compensated, for example, using the Tolles-Lawson model~\cite{canciani2021magnetic}. In contrast, time-varying disturbances are much harder to reduce. Still, they can be greatly reduced by placing the sensor array sufficiently far from the motor and electronics and, ideally, separating it from the AUV altogether~\cite{canciani2021magnetic, gnadt2022advanced}. 

\section{Conclusions}\label{sec:conclusions}

For constant one-dimensional motion, the derived CRLB expressions indicate the potential for a few-meter-long array with three contemporary quantum magnetometers to provide speed-estimation accuracy on par with that of an export-controlled DVL. This assumes that temporal variations in the magnetic field magnitude are accurately corrected, platform disturbances are negligible, and spatial field variations are sufficient. If temporal variations cannot be corrected, the bounds indicate that sensor noise must be reduced by approximately two orders of magnitude to reach the accuracy benchmark of an export-controlled DVL. Alternatively, the sensor array would need to be extended to lengths that may be impractical.

To conclude, the results suggest that a magnetometer array may complement, rather than replace, a DVL, providing an additional sensing modality to support inertial navigation. This could improve robustness and support covert operations. However, the analysis neglects platform disturbances, which in practice may be much larger than the sensor noise and limit performance. Future work should therefore investigate the effects of these disturbances and methods to compensate for them, and evaluate attainable velocity-estimation accuracy under more general vehicle motion.



\section*{Appendix}\label{app:Second-order}

For a unitary transform $U$, the following identity holds, 
\begin{equation}
    \Sigma^{-1}=U^H(U\Sigma U^H)^{-1}U
\end{equation}
where $(\cdot)^H$ denotes the Hermitian transpose operator.  The Fisher information is thus
\begin{equation}
\begin{split}
    \mathcal{I}_\theta=&\frac{\partial \mu(\theta)^\top}{\partial \theta}  \Sigma ^{-1}\frac{\partial \mu(\theta)}{\partial \theta}\\
    =&\frac{\partial \mu^\top(\theta)}{\partial \theta} U^H (U\Sigma U^H)^{-1}U \frac{\partial\mu(\theta)}{\partial \theta}.
\end{split}
\end{equation}

Consider the unitary transformation $U=I_N\otimes T_M$, where $T_M$ is the $M\times M$ discrete Fourier transform matrix, then,
\begin{equation}
\begin{split}
    (U\Sigma U^H)^{-1}
    =&\begin{bmatrix}
    \begin{smallmatrix}
        \left(MR_{\text{rw}}+\sigma_e^2I_{N}\right)^{-1} & 0  & \dots & 0\\
        0 & \frac{1}{\sigma_e^2}I_{N} &  \dots & 0\\
        \vdots  & \vdots & \ddots & \vdots &\\
        0 & 0 & \dots &  \frac{1}{\sigma_e^2}I_{N}
    \end{smallmatrix}
    \end{bmatrix},
\end{split}
\end{equation}
where assuming that \mbox{$ \sigma_e^2\ll M\sigma_w^2T_s^2$}, \begin{equation}
\begin{split}
\left(MR_{\text{rw}}+\sigma^2_e I_{N}\right)^{-1}=&\frac{1}{\sigma_e^2}\begin{bmatrix}
    \begin{smallmatrix}
        1 & 0 & \hdots & 0\\
        0 & 0 & \hdots & 0\\
        \vdots & \vdots & \ddots & \vdots\\
        0 & 0 & \hdots & 0
    \end{smallmatrix}
    \end{bmatrix}\\
    &+\frac{1}{M}\sum_{i=1}^{\infty}\left(-\frac{\sigma_e^2}{M}\right)^{i-1}(R_{rw}^+)^{i}.
\end{split}
\end{equation}
Here $(\cdot)^{+}$ is the pseudo-inverse.
Assuming that \mbox{$ \sigma_e^2\ll M\sigma_w^2T_s^2$} the matrix sum can be approximated as the first term, \begin{equation}
    \frac{1}{M}\sum_{i=1}^{\infty}\left(-\frac{\sigma_e^2}{M}\right)^{i-1}(R_{\text{rw}}^+)^{i}\approx \frac{1}{M}R_{rw}^{+}.
\end{equation} 
Thus 
\begin{equation}
\begin{split}
    \Sigma^{-1}\approx&
    \frac{1}{\sigma_e^2}I_{NM}+\frac{1}{M^2}R_{\text{rw}}^+\otimes1_M1_M^\top\\
    &-\frac{1}{M\sigma_e^2}\begin{bmatrix}
    \begin{smallmatrix}
        0 & 0 & \hdots & 0\\
        0 & 1 & \hdots & 0\\
        \vdots & \vdots & \ddots & \vdots\\
        0 & 0 & \hdots & 1
        \end{smallmatrix}
    \end{bmatrix}\otimes1_M1_M^\top.
\end{split}
\end{equation}
and the approximated Fisher information can then be calculated and further simplified under the assumption $ \sigma_e^2 \ll \sigma_w^2 T_s^2 $.

\bibliographystyle{IEEEtran}
\bibliography{refs}

\end{document}